\documentclass[12pt]{article}
\usepackage{epsfig}

\usepackage{graphicx}

\usepackage{epstopdf}
\catcode`\^^?=9 

\def\half{\mbox{\small{$\frac{1}{2}$}}}

\def\be{\begin{equation}}
\def\ee{\end{equation}}
\def\ba{\begin{eqnarray}}
\def\ea{\end{eqnarray}}

\def\ban{\begin{eqnarray*}}
\def\ean{\end{eqnarray*}}

\def\nref#1{(\ref{eq:#1})}
\def\mlab#1{\label{eq:#1}}

\begin{document}

\noindent
{ \large \bf Nonconservation of Energy and Loss of Determinism}
 \vspace{2mm}
 
 \noindent
 { \bf II. Colliding with an open set}
 \vspace{4mm}
  
  \noindent
 {\small {\bf David Atkinson,} 
 University of Groningen, 9712 GL   Groningen, The Netherlands}

\noindent
{\small {\bf Porter Johnson,} Illinois Institute of Technology, Chicago, IL 60616, U.S.A.}

 \vspace{7mm}
 
\noindent
{\bf Abstract}
An actual infinity of colliding balls can be in a configuration in which the laws of mechanics lead to logical inconsistency. It is argued that one should therefore limit the domain of these laws to a finite, or only a potentially infinite number of elements. With this restriction 
indeterminism, energy nonconservation and {\em creatio ex nihilo}   no longer occur. A numerical analysis of finite systems of colliding balls is given, and the asymptotic behaviour that corresponds to the potentially infinite system is inferred. 
\vspace{1mm}

\pagestyle{empty}

\section{Introduction}
Aristotle is known for his espousal of the idea of {\em potential infinity}, as opposed to {\em actual infinity}, which he generally denied to have any physical existence.\footnote{Except for his treatment of time past\cite{Aristotle}.} In his veto of the actually infinite, Aristotle was followed by the great majority of the mediaeval scholastics under the motto {\em `infinitum actu non datur'}. While mainstream mathematicians, in the wake of Georg Cantor, accept the coherence of  the idea of actually infinite, measurable sets, a rearguard of intuit\-ionalists deny their claims. In physics the conceptual use of infinite systems has always veered toward potential, rather than to actual infinity. The theoretical physicist simply {\em defines} an infinite system to be the conceptual result of allowing a finite system to grow without bound. In particular, all physical quantities, such as energy or momentum, that have a finite limit as the size of the system tends to infinity are decreed to be the `physical' quantities appertaining to the (potentially) infinite system. Quantities that diverge in this limit can also be interesting. As an example, consider the question whether a loss-free, infinite mechanical system can exhibit irreversible motion, rather than the Poincar\'e cycles that are the inevitable lot of a finite system.   
The standard way to answer this question is {\em not} to consider subsystems that are {\em ab initio} infinite in number. Rather, one considers $n$ subsystems, where $n$ is finite but variable, and calculates, or at any rate estimates the Poincar\'e recurrence time, $R$, as a function of $n$. The physicist is satisfied if she can show that this recurrence time  is unbounded; that is, for any finite time interval, $T$, there exists an integer, $N$, such that, for all $n>N$, it is the case that $R>T$.

Although physicists make exemplary use of the methods of infinitary mathematics, they may be said to be Aristotelians in their approach to infinity in physical systems. Not so certain contemporary philosophers of science.  Jon P\'erez Laraudogoitia\cite{Laraudogoitia} considered an actually infinite set of identical, colliding balls in which every individual collision is elastic and deterministic. Nevertheless, he found that neither 
energy nor momentum were conserved, and that indeterminism reigned. This is in stark contrast to the behaviour of a finite number, $n$, of such balls, for which both energy and momentum are conserved. The loss of energy of such a system of $n$ balls is zero for any finite $n$, and therefore it is zero also in the limit $n\longrightarrow \infty$.  Similarly momentum is conserved, and determinism prevails in this limit. Yet the actually infinite system of balls does not share these `physical' features.  Mathematically there is no paradox: there is simply a difference between the the infinite limit and the value at infinity --- we might call the phenomenon a discontinuity at infinity. Physically however we face a dilemma. Either we can embrace actual infinity and accept the loss of conservation and determinism, or we can, with Aristotle, seek to exclude actual infinity. 

\pagestyle{plain}
\pagestyle{myheadings}  \markright{}

In a previous paper\cite{Porter} (which we shall call I), where we built upon ideas that had been introduced by one of us\cite{Atkinson}, we analyzed a generalized version of Laraudogoitia's system of an actually infinite system of balls, in which however the masses of the balls were not all equal.  Our conclusion was that, also for this generalized system, the laws of mechanics (whether classical or relativistic) do {\em not} in general lead to conservation of energy-momentum, nor to determinism.  However, we showed that the actually infinite system {\em does} share with finite systems the property of time-reversal invariance. 

In Section 2 we consider the development that Alper and Bridger\cite{Alper1} gave to the   Zeno ball scenario, in which another ball approaches a point of accumulation of the actually infinite set of balls. We reject Alper and Bridger's conclusion that the ball must spontaneously disappear when it reaches this point of accumulation, arguing that the system involves a logically inconsistent set of properties. This logical inconsistency is removed  only by moving from the actual to a potential infinity of Zeno balls.  In Section 3 we approach the potentially infinite system by considering  a steadily increasing, but finite number of Zeno balls. When the Zeno balls are all identical, we find that the  ball of Alper and Bridger  either bounces back from the Zeno balls, or comes to rest, depending on the mass of the new ball. For masses that decrease geometrically, we give  the results of numerical calculations in which it is shown that qualitatively similar behaviour obtains, except that, when the new ball is sufficiently massive, it merely slows down, i.e. it moves with a  reduced positive velocity after all collisions have taken place. In Section 4 we conclude that, since the actually infinite system of Zeno balls encapsulates a logical inconsistency, any claim based on its analysis, for example that the laws of mechanics imply  neither conservation of energy nor determinism, is a {\em non sequitur}.

\section{Actual infinity of Zeno balls}
An infinite number of  identical, stationary point masses (Zeno balls) are placed at the Zeno points  1, $\frac{1}{2}$, $\frac{1}{4}$, $\frac{1}{8}\ldots $ on a straight line. A further ball, which we shall call the AB ball, is identical to the others and is situated on the line to the left of the origin, 0. It moves with constant speed towards the 
origin (see Fig. 1, where * denotes the origin). 

\begin{picture}(0,45)(-23,-30)
\put(-23.4,0){$\rightarrow$}
\put(100,0){\sf o}
\put(50,0){\sf o}
\put(25,0){\sf o}
\put(12.5,0){\sf o}
\put(6.25,0){\sf o}
\put(-25,0){\sf o}
\put(3,0.9){\tiny ...}
\put(0,-1.22){*}
\put(20,-15){\sf Figure 1. AB ball and Zeno balls}
\put(0,-6){\bf \small 0}
\put(100.2,-6){\bf \small 1}
\end{picture}

\vspace{-8mm}

\noindent
There is no ball at 0, but the origin is a point of accumulation of the locations of the Zeno balls. If the AB ball were to collide with a Zeno ball, it would come to rest, thereby imparting all its energy to the Zeno ball, which would  move off with the speed that the AB ball originally had. However, there is no Zeno ball with which it could collide. For suppose, {\em per impossibile,} that it did collide with one of the Zeno balls. Then it should first have collided with that Zeno ball's immediate left-hand neighbour, and this would have brought it to rest, making the posited collision impossible. Thus the AB ball can collide with none of the Zeno balls. In the absence of any forces other than those arising from collision, the AB ball must continue in its state of constant motion (Newton's first law), thus arriving in a finite time at the Zeno point 1. But this is also impossible, since an infinite number of Zeno balls should have blocked its way. 

Such is the scenario sketched by Alper and Bridger\cite{Alper1}, and these authors conclude that the AB ball must simply cease to exist when it arrives at the origin, since it can neither be stopped there, nor can it progress further, nor indeed can it be anywhere else. Not only has energy disappeared without trace, as in P\'erez Laraudogoitia's related problem, but a massive ball has vanished! Alper and Bridger further speculate about a time-reversed scenario in which a ball suddenly pops into existence at the point of accumulation of a stationary line of Zeno balls and then moves away from them. This amounts to {\em creatio ex nihilo} of inertial mass and kinetic energy in one fell swoop.    

It is one thing to demonstrate (as we did in paper I) that time-reversal invariance is a mathematical property of the (actually) infinite system of equations, and quite another `to move', as Angel\cite{Angel} remarks, ` from the mathematical availability to the metaphysical assumption' that the spontaneous creation a ball is a serious possibility. However, to claim. as he does,  that this might offend `a basic plausibility assumption' is stretching intuition well beyond breaking point. Fortunately, a logical attack is possible, rather than one grounded on vague plausibility.

Alper and Bridger's system constitutes a logical contradiction. To be precise the following conditions are inconsistent with one another:\footnote{This, and other inconsistent systems that are isomorphic to it, have been considered by Peijnenburg and Atkinson\cite{Jeanne}.} 
 \begin{enumerate}
 \item
Stationary balls of unit mass and zero size -- mass points -- are situated at the Zeno points.  
 \item 
A moving ball of unit mass and zero size travels at constant speed, reaching 0 from the left. 
  \item 
 When the moving mass point occupies the same position as a stationary mass point, it comes to rest, otherwise it continues in its state of constant motion.  
 \item 
 The moving ball comes to rest before reaching the point 1. 
 \end{enumerate}
While it is formally possible in classical logic to deduce anything from a contradiction {\em (ex contradictione quodlibet),}  
 of course to say anything significant about a physical system (however idealized), one should start from a noncontradictory set of statements. Moreover, Alper and Bridger's statements are doubly suspect, for they ask us to believe that there is no trouble until the AB ball reaches the origin, but that the system becomes paradoxical at the moment that the ball reaches that point. However, that is a misreading of the situation: the system as described by the four above conditions is inconsistent {\em tout court}, not simply at one particular time.

A straightforward way to avoid the inconsistency is to back the Aristotelian option and ban an actual infinity of balls. As we will see in the next section, in the case that the number of Zeno balls is only potentially infinite, we can easily answer the question as to what happens to the AB ball.

\section{Potential infinity of Zeno balls}
To approach potential infinity we analyze the case of a finite number, $n$,  of identical Zeno balls, each of unit mass, and we will consider the limit as the variable $n$ grows beyond any bound. At the same time we treat  the mass of the AB ball, $m_{AB}$, as a variable. If a ball of mass $M$, moving at velocity $v$, strikes a stationary ball of mass $m$, its velocity changes to $v^{new}$,  where
\ba
v^{new}&=&\frac{M-m}{M+m}\, v\,.
\mlab{basic}
\ea
Consider the first collision of the AB ball: its initial velocity is $v=1$, and if $m_{AB}<1$, it will suffer no further collisions. Its final velocity will be in fact 
\be
v_{AB}=\frac{m_{AB}-1}{m_{AB}+1}\,,\mlab{ABball}
\ee
which is negative, since $m_{AB}<1$. The formula is also correct when $m_{AB}=1$, of course, since then the AB ball is brought to rest by collision with an equally massive Zeno ball.

However, in the case $m_{AB}>1$ the AB ball suffers precisely $n+1$ collisions. The first collision reduces its velocity to the value \nref{ABball}, which is now positive, and a first wave of collisions progresses through the set of Zeno balls, leaving them finally all at rest except for the last one (i.e.  Zeno ball number 0), which escapes without further collision. The second collision of the AB ball with Zeno ball number $n$ will reduce its velocity to the square of the right-hand side of \nref{ABball}, the third collision to the cube of the right-hand side of \nref{ABball}, and so on. The 
$(n+1)$st (and final) collision of the AB ball with Zeno ball number $n$ reduces its velocity to 
\be
v_{AB}=\left(\frac{m_{AB}-1}{m_{AB}+1}\right)^{n+1} \,.\mlab{ABballplus}
\ee
It should be noted that $v_{AB}$ is positive for any finite $n$: it is only in the limit of an infinite number of identical Zeno balls that the AB ball comes to rest, and moreover it does so in a finite time. Thus the AB ball  bounces off an infinite set of identical Zeno balls only when the mass of the AB ball is less than that of a Zeno ball. When its mass is equal to, or greater than that of a Zeno ball, the AB ball   
comes to rest. We can also look at 
 a system in which the masses decrease in geometrical progression. By allowing the number and the rate of decrease to vary, we shall gain numerical insight into the nature of the double limit in which there are infinitely many balls of identical mass.  
Suppose that the AB ball has unit mass, and that it moves with unit velocity towards the right, as in Figure 1, but there are $n+1$ stationary Zeno balls only. The masses of these Zeno balls are 
\[
m_p=\mu^p\hspace{8mm}{\rm with}\hspace{8mm}p=0,1,2,\ldots,n\,.
\]
\vspace*{5mm}

\begin{figure}[h]
\begin{picture}(0,60)(0,5)
\put(8,13){\mbox{\epsfig{file=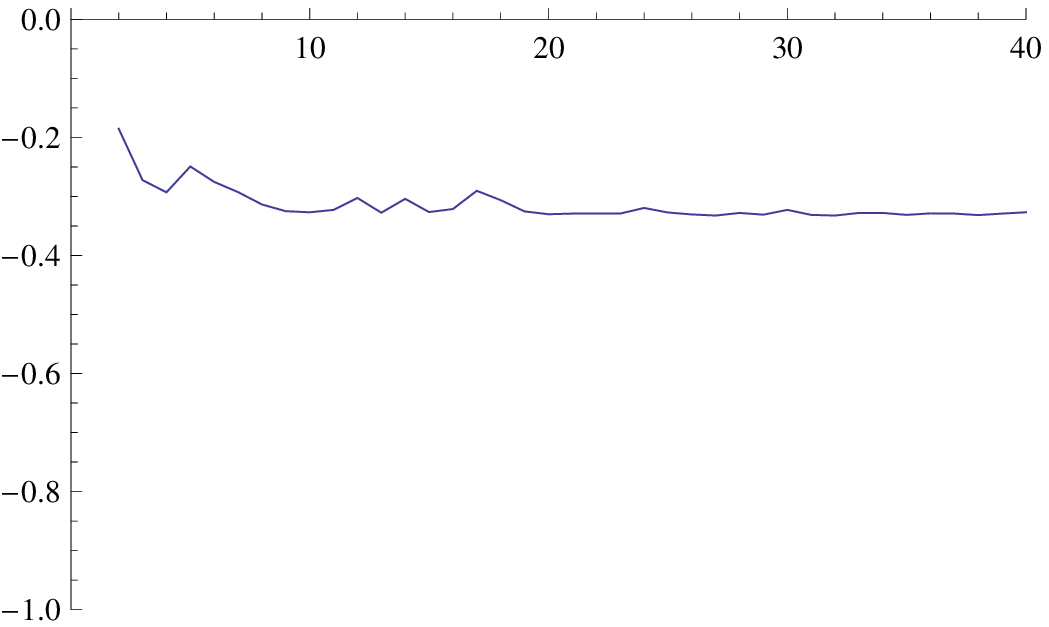,height=5cm}}}
\put(11,65){$v_{AB}$}
\put(40,66){Number of Zeno balls}
\put(10,4){Figure 2. Final velocity of AB ball:  $m_{AB}=1$ and  $\mu =0.5$}
\end{picture}
\end{figure}

\begin{figure}[t]
\begin{picture}(0,0)(0,70)
\put(8,13){\mbox{\epsfig{file=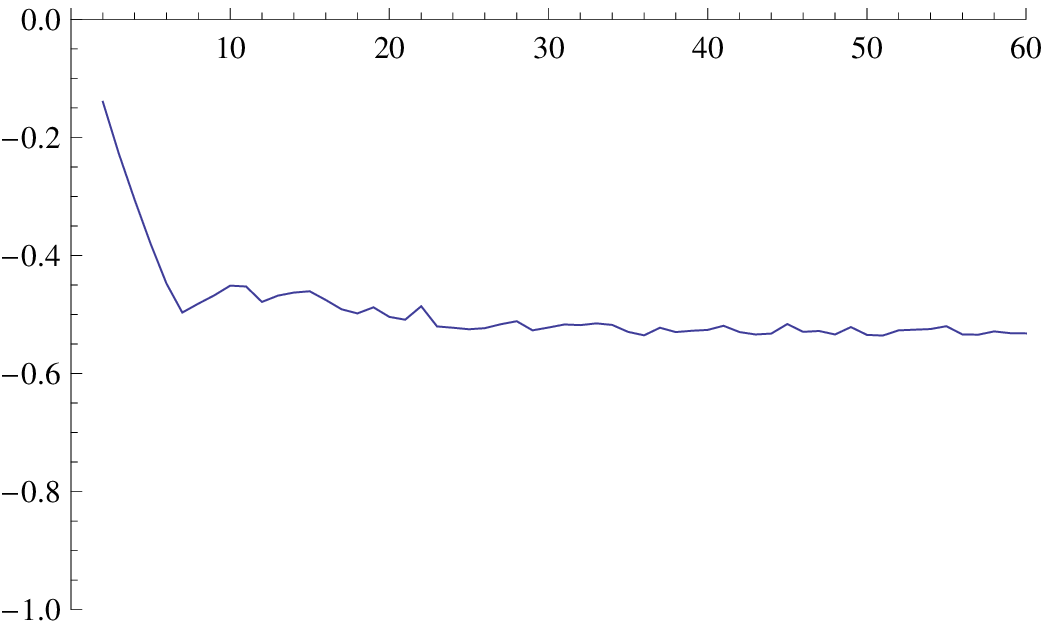,height=5cm}}}
\put(11,65){$v_{AB}$}
\put(40,66){Number of Zeno balls}
\put(10,4){Figure 3. Final velocity of AB ball:  $m_{AB}=1$ and  $\mu =0.7$}
\end{picture}
\end{figure}

\vspace*{65mm}
In Figures 2 -- 4 the final velocity of the AB ball, $v_{AB}$, is shown for  $\mu =0.5, \, 0.7$ and $0.95$, as a function of the number of Zeno balls. As can be seen from these graphs, the qualitative behaviour of the AB ball is that it rebounds from the set of Zeno balls, but with reduced speed. 
This bears some resemblance to what would happen in an {\em inelastic} collision of the AB ball with a single ball of mass greater than unity. It can be seen that the final speed is greater for 
larger values of $\mu$, and this makes good sense, since the total mass of all the 

 \begin{figure}[b]
\begin{picture}(0,50)(0,30)
\put(8,13){\mbox{\epsfig{file=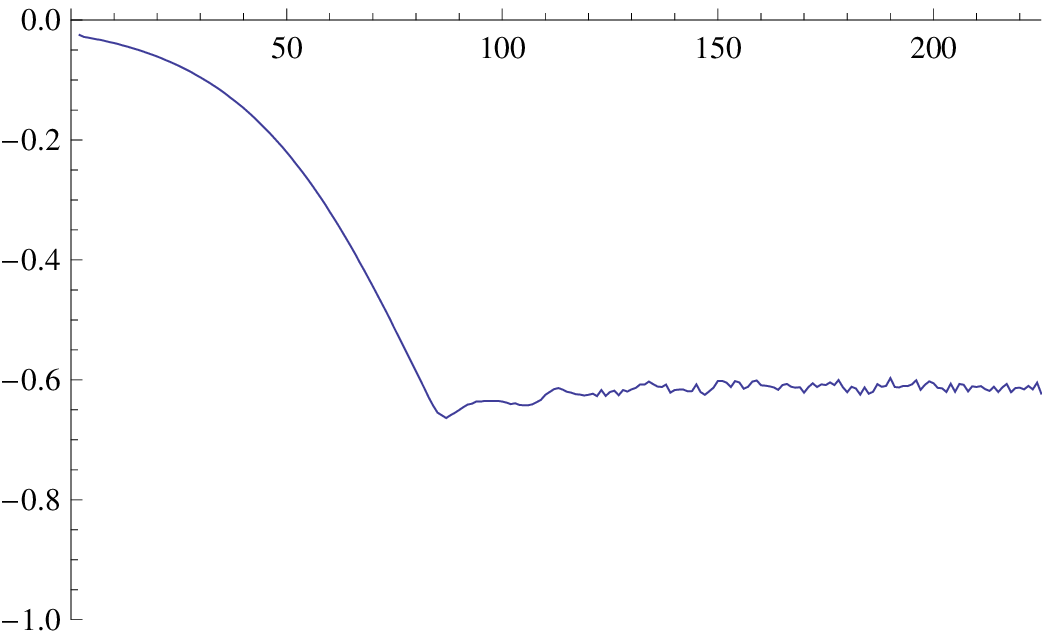,height=5cm}}}
\put(11,65){$v_{AB}$}
\put(40,66){Number of Zeno balls}
\put(10,4){Figure 4. Final velocity of AB ball:  $m_{AB}=1$ and  $\mu =0.95$}
\end{picture}
\end{figure}

\vspace*{65mm}
\noindent
Zeno balls is an increasing function of $\mu$.  A point of interest is that the dependence on $n$ is not entirely smooth. The total number of collisions is a rapidly increasing function of the number of balls. The wobbles are caused by the fact 
that the system is discrete: even if there is not strictly speaking a limiting value for 
$v_{AB}$ as $n\longrightarrow\infty$, a suitably smoothed version, such as the Cesar\`o mean, probably does have a limit. This limit can be approximately given for the various values of $\mu$ from the estimated asymptote to the curves. 
 Figures 5 --- 6 show the final velocity of the AB ball  for $\mu =\half$ as a function of the AB mass.   In Figure 5 the results are shown for a number of values up   
 \begin{figure}[t]
\begin{picture}(0,0)(0,65)
\put(8,18){\mbox{\epsfig{file=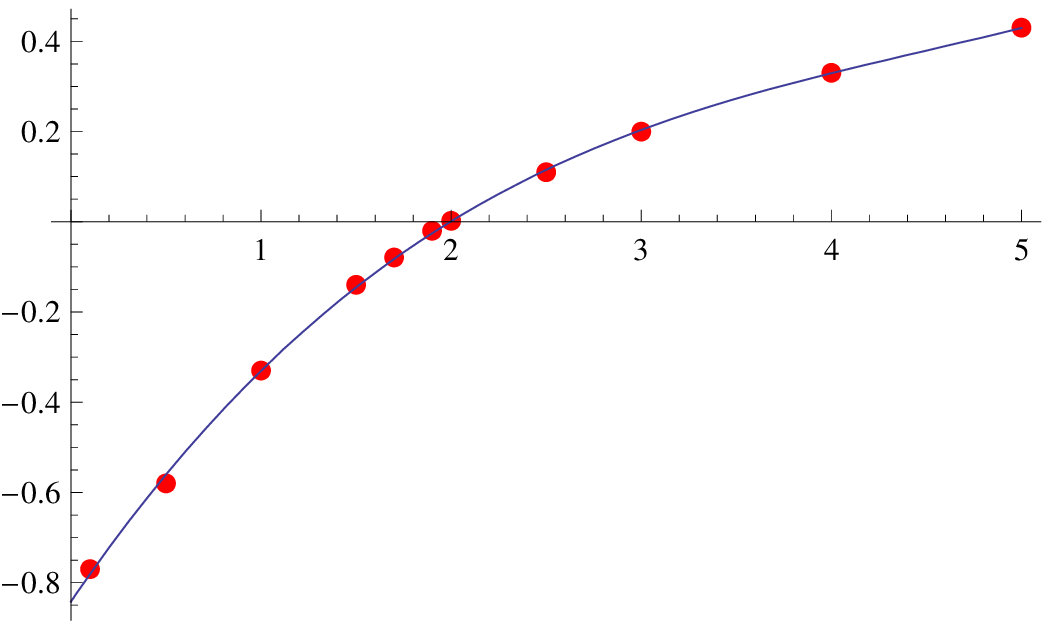,height=5cm}}}
\put(11,70){$v_{AB}$}
\put(95,50){$m_{AB}$}
\put(10,5){Figure 5. Final velocity of AB ball as a function of its mass:  $\mu =\half $}
\end{picture}
\end{figure} 

\begin{figure}[b]
\begin{picture}(0,0)(0,10)
\put(14.3,19.5){\mbox{\epsfig{file=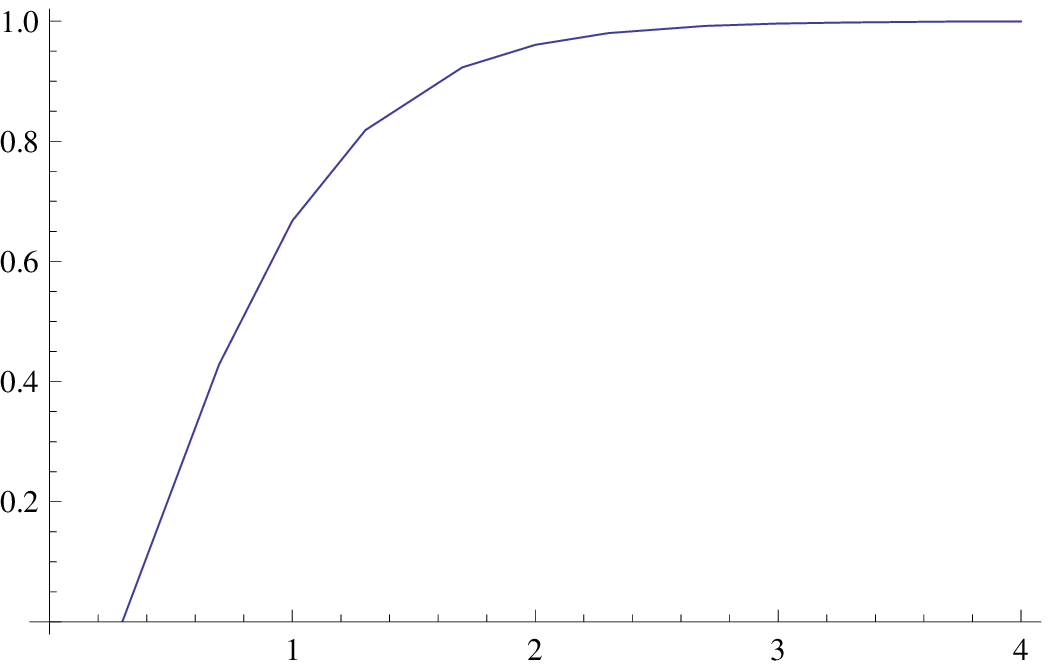,height=5cm}}}
\put(13,72){$v_{AB}$}
\put(95,21){$m_{AB}$}
\put(33.8,18.5){\tiny \sf 10}
\put(51.7,18.5){\tiny  \sf 10}
\put(70.1,18.5){\tiny  \sf 10}
\put(88.6,18.5){\tiny  \sf 10}
\put(10,5){Figure 6. As in the previous figure, but with a logarithmic mass-scale}
\end{picture}
\end{figure}

\newpage

\noindent
to $m_{AB}=5$, and a quadratic fit  to the computed points  has 
been  made.  Figure 6 gives the final velocity up to $m_{AB} =10000$, against a logarithmic scale.  It can be seen that the AB ball comes to rest when $\mu$ is approximately 2, which is the total mass of all the Zeno balls together, since 
$m_n=2^{-n}$. 
This is what would happen if all the Zeno balls were concentrated into one ball of mass 2. Actually, the AB ball is brought to rest when $m_{AB}\approx 1.96$, and this reflects the fact that there is some slight motion of the Zeno balls relative to one another, after they have all suffered their final collisions.

\section{Ontology of the infinite}
Laraudogoitia's system of Zeno balls appears to obey Newton's laws of motion (Laraudogoitia, Earman and Norton argue that the system does, while Alper and Bridger assert that it does not really obey these laws\cite{Alper1, Norton1, Alper2, Alper3, Alper4,Laraudogoitia2}). It also exhibits time-reversal invariance (this is clear locally, and see I for a formal proof of the existence of one solution that is the precise time-inverse of the forward solution of the equations of motion\footnote{When an equation in mathematical physics has more than one solution, there may be a good physical reason to exclude some of those solutions. For example, the distance, $d$, that a body falls in time, $t$, is given by $d=\frac{1}{2}gt^2$, where $g\approx $ 10 m/s$^2$ is the acceleration due to gravity. If we fill in 5 m for $d$, we obtain two solutions for $t$, namely 1 s or $-1$ s. The negative option is deemed `unphysical', and nobody worries about the existence of this extra solution. In the case of an actual infinity of Zeno balls we {\em could} outlaw the solutions that violate energy conservation. Even the original scenario of Laraudogoitia can be tamed by choosing the solution in which a ball pops into existence at the opportune moment, just in time to carry away the energy-momentum that would otherwise be lost. This `popping into existence' is just the time reverse of the Alper-Bridger scenario, in which a ball suddenly disappears. This is of course an {\em ad hoc} way of saving the conservation laws, and we prefer to rule out actual infinity as a radical extirpation of disease, but the above palliative is open to those who accept the Alper-Bridger viewpoint.}). The system is therefore wildly indeterministic, since spontaneous motion can arise at any time, with creation of energy out of nothing. The question arises however as to whether infinite systems of elements, like the Zeno balls, properly belong to the domain of Newton's mechanics. We are familar with useful laws that are valid in restricted domains, indeed most laws have some limitation as to the kinds of situations that they can reasonably be expected to handle. For example, Newton's laws break down when velocities are not very small compared with that of light, but this escape route is blocked, for we showed in I that loss of energy-momentum and indeterminism reign also in the special relativistic domain. Einstein cannot rescue Newton. 

Three routes at least are open:
\begin{enumerate}
\item 
Accept that the laws of mechanics entail neither energy conservation nor determinism. This is the option of many philosophers of science, who do not flinch from upsetting their physicist confr\`eres\cite{Laraudogoitia}, \cite{Norton1}.
\item 
Decree that actually infinite systems lie outside the domain of mechanics --- whether classical or relativistic. This would be the choice of the hard-headed mechanic who thinks he `knows' that energy cannot be created out of nothing, and who is therefore not interested in a theory that envisages such an outr\'e possibility. 
\item 
Show that there is a logical impediment to the application of the laws of mechanics to systems involving an actual infinity of elements. The reason for the restriction of the domain to the finite, or the potentially infinite only, would then not  be merely a matter of prejudice, but would be forced by logic.
\end{enumerate}
The third option is the one that we espouse. 

If one were to include the system involving an actual infinity of  Zeno balls as a card-carrying member in the domain of the laws of mechanics, one would also have to admit these balls together with the AB ball. In particular there would be no valid {\em a priori} objection to admitting the applicability of the laws to the configuration in which the AB ball arrives at the point of accumulation of the  positions of the Zeno balls. We have argued that this configuration leads to a logical contradiction, so it does not make sense to try to apply the laws to it. To remove the absurdity we stipulate that systems involving an actual infinity of elements do not lie within the domain of Newtonian, or of Einsteinian mechanics. This is not the same as saying that the conservation laws break down if the system is actually infinite: rather mechanics itself does not make sense. The claims that the laws of mechanics do not lead to the conservation of energy and momentum may thus no longer be maintained, and rampant indeterminism is seen as a chimera.  

It would be nice if the banning of actually infinity were an effective remedy for the manifold ills besetting energy-conservation and determinism. By excluding actual infinity in space and in time one certainly escapes some of Earman's\cite{Earman}  challenges; and by forbidding the actual infinity of higher-order spatial derivatives one avoids impalement on Norton's dome\cite{Nortondome}. Nevertheless, many lacunae remain in the comfortable, illusory world of deterministic mechanics; and proscribing the actually infinite is merely a necessary, not a sufficient stratagem to stop them all.


\section*{Appendix}
In this appendix we explain in outline the architecture of the computer programme that was written to determine the final velocity of the AB ball, and the final velocities of all the Zeno balls. The mass of the AB ball is $m_{AB}$, and the masses of the Zeno balls are $m_n,\, n=0\ldots N$, where $N$ is finite. It is convenient to call the AB ball the $(N+1)$st ball, setting  formally $m_{N+1}\equiv m_{AB}$. The AB ball is situated always to the left of the Zeno ball of mass $m_N$, with which it repeatedly collides, and the Zeno balls also collide with each other. 

Consider an intermediate time at which some but not all of these collisions have taken place. The intermediate configuration is specified by the positions, $x_n$, and the velocities, $v_n$ of all the balls. The index $n$ run over the values $ n=0\ldots N+1$, where 
$x_{N+1}$ and $v_{N+1}$ stand for the current position and velocity, respectively, of the AB ball.  If the velocities of approach of all the balls are negative (or zero), no further collisions will occur, and the programme terminates. If one or more of the velocities of approach are positive, however, the time interval, $\Delta T$, is calculated for the earliest collision, and the index, $k$, is noted, such that the two balls that come into contact are those with masses $m_{k-1}$ and $m_k$. The changes in the velocities of these balls, as a result of the collision, are then computed. The iterative process is as follows:

\begin{itemize}

\item

For each integer $n$, in the closed interval $[1, N+1]$, determine
whether the pair of adjacent masses $(m_{n-1}, m_n)$ are approaching
one another.

\item

If the masses are approaching one another, determine the projected time
$\delta t_n$ at which they would collide.

\item

 Determine the time for the next collision in the system, $\Delta T$, 
as the {\sf minimum} of the quantities $\{ \delta t_n \}$, and   note the index $k$ that indicates which pair of particles will first collide. If none of the pairs are approaching one another,  terminate the programme. 

\item
If there are subsequent collisions, update the time
and the positions of all the masses:
\ban
t &\rightarrow& t + \Delta T \\
x_n &\rightarrow& x_n + v_n \, \Delta T
\ean

\item

Change the velocities of the 
pair  that collide:

\ban
v_{k-1}^{new}&=&\frac{2 \, m_k}{m_{k-1} + m_k} \, v_{k-1} +
\frac{m_{k-1} - m_k}{m_{k-1}+m_k} \, v_k\,,  \\
v_{k}^{new}&= & v_k - v_{k-1} + v_{k-1}^{new}\,,
\\   v_{k-1}^{new}&\rightarrow& v_{k-1}
\\  v_{k}^{new}&\rightarrow& v_{k}
\ean
\end{itemize}

\noindent
This process must be iterated until it stops. The programme can be further represented by the 
flow chart given in Table 1. 

\newpage

\begin{center}
\begin{picture}(125,140) (-10,20)

\put(7,120){\framebox(60,15)}
\put(12,130){Initial configuration of balls}
\put(16,125){$\{x_n, v_n \}$; set time $t = 0$}
\put(35,115){$\downarrow$}

\put(7,95){\framebox(60,15)}
\put(10,105){Configuration of balls}
\put(14,100){at time $t$: $\{x_n, v_n \}$ }
\put(35,90){$\downarrow$}

\put(7,70){\framebox(60,15)}
\put(13,80){Locate next collision}
\put(14,76){between balls $k-1$ and $k$}
 \put(14,72){after time interval $\Delta T$}
\put(35,64){$\downarrow$}

\put(7,41){\framebox(60,20)}
\put(10,56){Update}
\put(18,52){time: $t \rightarrow t + \Delta T$}
\put(10,48){positions: $x_n \rightarrow x_n + v_n \, \Delta T$}
\put(8,44){ velocities: $v_{k-1}$ and $v_{k}$}
\put(35,36){$\downarrow$}

\put(7,25){\framebox(67,8)}
\put(8,28){Velocities of approach all negative?}
\put(80,27){$\rightarrow$}

\put(92,25){\framebox(10,8)}
\put(94,28){No}

\put(35,20){$\downarrow$}
\put(22,10){\framebox(25,8)}
\put(25,13){Yes: Stop}

\put(95,43){$\uparrow$}
\put(95,53){$\uparrow$}
\put(95,63){$\uparrow$}
\put(95,73){$\uparrow$}
\put(95,83){$\uparrow$}
\put(95,93){$\uparrow$}

\put(85,100){\framebox(25,8)}
\put(91,103){Iterate}

\put(75,102){$\leftarrow$}

\put(25,-10){ Table 1. Flow Chart for Collisional Process}

\end{picture}
\end{center}

\end{document}